\documentstyle[twoside,fleqn,espcrc2,epsfig]{article}

\newtheorem{theorem}{Theorem}[section]
\newcommand{\MathC}{\mbox{\sf \hspace*{0.25em}\rule[0.05em]{0.035em}{0.59em}\hspace*{-0.285em}C}}

\newcommand{\MathR}{\mbox{\sf I\hspace*{-0.12em}R}}

\newcommand{\eqnref}[1]{\mbox{(\ref{#1})}}

\newcommand{\AmS}{{\protect\the\textfont2
  A\kern-.1667em\lower.5ex\hbox{M}\kern-.125emS}}

\title{Linear systems solvers -- recent developments and implications 
for lattice computations}

\author{A. Frommer\address{Fachbereich Mathematik, Universit\"at Wuppertal, \\
     42097 Wuppertal, Germany}}
       
\begin{document}

\begin{abstract}
We review the numerical analysis' understanding of Krylov subspace methods 
for solving (non-hermitian) systems of equations and discuss its implications
for lattice gauge theory computations using the example of the Wilson fermion
matrix. Our thesis is that mature methods like QMR, BiCGStab or restarted
GMRES are close to optimal for the Wilson fermion matrix. Consequently,
preconditioning appears to be the crucial issue for further improvements.
\end{abstract}

\maketitle

\section{KRYLOV SUBSPACE METHODS}

Given a linear system of equations
\begin{equation} \label{basic_eq}
Mx = b 
\end{equation}
with $M \in \MathC^{n \times n}$ being non-singular, the class of
Krylov subspace iterative methods for solving \eqnref{basic_eq}
is characterized by the following generic template
\begin{tabbing}
for \= for \= for \= \kill
choose initial guess $x^0$, set $r^0 = b - Mx^0$ \\
 for $m=1,2,\ldots$ until convergence \\
 \> compute iterate $x^m$ of the form \\
 \> \> $x^m = x^0 + q_{m-1}(M)r^0$
\end{tabbing} 
Here, $q_{m-1}$ is a polynomial of degree $\leq m-1$. For the residual 
$r^m = b - Mx^m$ we
therefore get
\begin{equation} \label{res_eq}
r^m = p_m(M)r^0,
\end{equation}
where $p_m$ is the polynomial $p_m(t) = 1 - tq_{m-1}(t)$. In an algorithmic description 
of virtually any Krylov subspace method, the polynomials $q_{m-1}$ or $p_m$ are not
explicitly present, but they are crucial to a theoretical analysis of the method. 
Moreover, the relation \eqnref{res_eq} is also the key to understanding the condition
(`difficulty') of the linear system to be solved and we start by discussing this point.

\subsection{Condition}
Assume that $M$ is diagonaizable, i.e.\ we have a decomposition of the form
\[  
M = V \Lambda V^{-1}
\]
where $\Lambda$ is diagonal with its diagonal containing the eigenvalues, and $V$ is 
the corresponding matrix of (right) eigenvectors. Denoting the spectrum
of $M$ by $\sigma(M)$, 
from \eqnref{res_eq} we now get $r^m = V p_m(\Lambda) V^{-1} r^0$ and therefore
\begin{equation} \label{bound_eq}
\|r^m\| \leq \|V \| \cdot \|V^{-1}\| \cdot \|r^0\| \cdot \|p_m(\Lambda)\|.
\end{equation} 
Since $p_m(\Lambda)$ is diagonal we have
\[
\|p_m(\Lambda)\| = \max_{\lambda \in \sigma(M)} |p_m(\lambda)| .
\]
Considering $\|V\|, \|V^{-1}\|$ and $\|r^0\|$ as constants, the best
bound any Krylov subspace method can achieve in \eqnref{bound_eq} is the one obtained
for the polynomial which minimizes $\|p_m(\Lambda)\|$. In this sense, the quantities
\begin{equation}  \label{cm_eq}
c_m := \min_{{\deg(p_m) \leq m, \atop p_m(0) = 1}} \max_{\lambda \in \sigma(M)} |p_m(\Lambda)|, \;
 m=0,1,\ldots
\end{equation}
represent a measure of the {\em condition} of the system \eqnref{basic_eq}, since no Krylov
subspace method can achieve a better bound in \eqnref{res_eq} than the one which replaces
 $\|p_m(\Lambda)\|$ by $c_m$. Finding the optimal polynomial in \eqnref{cm_eq} is
a complex approximation problem for which solutions are known only in special cases. 
However, it is clear that due to the restriction $p_m(0) = 1$ the numbers $c_m$ will
tend to zero only slowly if there are many eigenvalues close to 0, particularly if 
they are distributed quite evenly around $0$.

\subsection{Optimal methods}
A Krylov subspace method is feasible algorithmically if it requires only a finite
amount of ressources like storage and computer time. We express this fact by saying that
the method can be implemented using {\em short recurrencies}, meaning that all quantities
needed at iteration $m$ can be computed from those of a small number of previous
iterations. Note that each Krylov subspace method will require at least one multiplication with
$M$ per iteration to account for the fact that the degree of the polynomial $p_m$ 
will increase as the iteration proceeds. The following theorem \cite{FM84} 
shows that optimality {\em and} short recurrencies can only be achieved for a restricted
class of matrices.

\begin{theorem}
A Krylov subspace method which achieves optimality, i.e.
\begin{equation} \label{opt_eq}
\|r^m\| = \min_{{\deg(p_m) \leq m, \atop p_m(0) = 1}} \|p(M)r^0\|
\end{equation}
for every initial residual $r^0$ and which can be implemented using short
recurrencies exists only if $M$ is of the form
\[
M = e^{i\Theta}(T + i \sigma I), \; \mbox{ where } T = T^{\dagger} \mbox{ and }
\sigma, \Theta \in \MathR.
\]
\end{theorem}
This theorem also holds if $\| \cdot \|$ is replaced by an energy norm of
the form $\|x\|_H = x^{\dagger}Hx$ with $H \in \MathC^{n \times n}$ hermitian and positive
definite. 
For $M$ hermitian and positive definite the CG method achieves optimality
in the energy norm with $H = M$. For $M$ hermitian (but possibly indefinite), 
MINRES \cite{PS75} is optimal in the Euclidian norm. 
The paper \cite{Fr90} gives algorithmic descriptions for optimal methods
in the other cases of Theorem~1.1.
Note that the above theorem includes matrices
of the form $\sigma I + S$ with $S^{\dagger} = - S$ (take $\Theta = - \pi/4$),
arising for staggered fermions.

\section{NON-HERMITIAN SYSTEMS}
The last 10 years have seen tremendous progress in Krylov subspace methods
for solving linear systems which, like the Wilson fermion matrix,
do not fall into the category covered by Theorem~1.1. See \cite{FGN91,Sa96}
for an overview. 
For simplicity, such systems will just be termed `non-hermitian' in the sequel.
In these cases 
one must find an adequate compromise between the quality of the Krylov subspace
method to use and the ressources required by the method. 

The first method
of this kind is the BiCG method \cite{Fl75}. Here, an additional shadow 
residual $\tilde{r}$ 
is selected and the $m$-th iterate $x^m$ is defined by the Galerkin condition
\[
(r^m)^{\dagger}\tilde{p}_m(M)\tilde{r} = 0
\]
for all polynomials $\tilde{p}_m$ of degree $\leq m$. In case that $M$ is 
hermitian positive definite and $\tilde{r} = r^0$ the method reduces
to the CG method. BiCG needs two matrix multiplies (one with $M$ and one 
with $M^{\dagger}$) per iteration and the residuals typically undergo quite
large variations. Moreover, there are situations where the method breaks down
(due to division by zero) without having reached a solution. Although exact
breakdowns do rarely occur in practice, near breakdowns severely affect the
numerical stability.

\subsection{QMR}

QMR, the quasi minimal residual method of \cite{FN91}, can be regarded as one
way to make BiCG more reliable. As BiCG it is based upon the non-symmetric
Lanczos process to compute an appropriate basis $ v_1, \ldots, v_m $ of the 
Krylov subspace $ K_m (M, r^0) = \{ p_l (M) r^0, \, \deg p_l \leq m-1 \} $. The $m$-th 
residual $ r^m$ is characterized by the coefficient vector $ ( \alpha_1, \ldots, 
\alpha_m) $ in $ r^m = \sum_{i = 1}^m \alpha_i v_i $ having minimal norm subject
to the condition $ r^m = p_m (M) r^0, \, \deg p_m \leq m, \, p_m (0) = 1 $. If 
the Lanczos vectors $ v_1, \ldots, v_m $ were orthogonal this would imply that 
$ r^m $ is minimal. Since for non-hermitian matrices the Lanczos vectors are not
orthogonal, minimizing the coefficient vector merely implies a `quasi' 
minimality of $ r^m $ whence the name QMR. QMR eliminates one source of 
breakdowns present in BiCG. Moreover, using a look-ahead strategy in the 
non-symmetric Lanczos process, almost all other (exact or near) breakdowns are 
also avoided at the price of extra storage. All these features are implemented 
in QMRPACK which is available from netlib. As in BiCG each iteration costs one 
multiply with $ M $ and one with $ M^\dagger $. The quite smooth 
convergence of QMR is also justified by the theoretical analysis.

\subsection{$J$-hermitian matrices}
A matrix $ M $ is said to be {\em $ J$-hermitian} if there exists a matrix $J$ such 
that
\[
M J = J M^\dagger \ .
\]
In this particular case, the non-symmetric Lanczos process can be made less 
costly, since through the right choice of the `shadow residual' $ \widetilde{r}
$ all multiplications with $ M^\dagger$ can be replaced by multiplications with
$ J $ \cite{Fr94}.
Consequently, BiCG and QMR require only one multiply with $ M $ and one with 
$ J $ in each iteration. For the Wilson fermion matrix we have $ J = 
\gamma_5 $ and thus multiplies with $ J $ are by far more cheaper than with 
$ M $. Exploiting the $\gamma_5$-symmetry thus makes QMR (and BiCG) competitive
to the other methods discussed in this section, see \cite{Fo96,Fretal95}.
At the time of writing this 
article, including the $J$-hermitian case into QMRPACK was under 
preparation \cite{FN96} but not yet completed.

\subsection{BiCGStab}

The BiCGStab \cite{vV92} method is another way to stabilize BiCG. Here, 
multiplies with $ M^\dagger$ are replaced by multiplies with $M$ such that an
additional one-dimensional minimization process is performed during each
iteration.

All computational effort, in particular, all matrix multiplies is spent working
on the iterates of the system to solve. Typically, BiCGStab produces less
varying residuals than BiCG, although the same sources for breakdowns are still
present. BiCGStab is quite easy to implement `from scratch'. Some
variations are described in \cite{Gu94,SF93}

\subsection{Restarted GMRES}

In contrast to the Lanzcos process, 
the Arnoldi process computes an {\em orthogonal} basis of $ K_m (M, r^0) $
for a general non-hermitian matrix $M$. From the Arnoldi basis it is possible 
to calculate an optimal iterate $ x^m $ (such that $ r^m $ satisfies \eqnref{opt_eq} )
by solving a small least squares problem.

The resulting method is called GMRES, the generalized minimal residual 
method \cite{SS86}.
However, the Arnoldi process does not rely on short recurrencies requiring
$m$ vectors of storage and $ O (m^2)$ inner products to be computed.

One therefore has to stop GMRES after a certain number ($k$,  say) of
iterations and restart the process with the current iterate $x^k$ as a new
initial guess. The resulting method is termed restarted GMRES or GMRES($k$).
For $ k = 1 $, a restart is done after every iteration. Hence, GMRES(1) is
identical to the familiar MR method \cite{EES83},
where the iterate $x^{m+1}$ is obtained
by minimizing $ r^{m+1} (t) = b - M (x^m + t r^m) $ with respect to $t \in \MathC$.
There are situations
where GMRES($k$) stagnates without reaching a solution, even for large restart
values $k$, but if all eigenvalues of $M$ lie in the right half plane GMRES($k$)
is known to converge for all $k$ \cite{Sa96,SS86}.

\section{PRECONDITIONING}

We have seen in Section~1 that the eigenvalue distribution of $M$ determines
a bound on the maximal speed of any Krylov subspace method for $M$. 
Once we have a method
which is close to optimal, the only way of getting further improvement is to 
change the matrix $M$ to one for which the eigenvalue distribution is more 
favorable. This is precisely the purpose of preconditioning where the original
system $Mx = b$ is changed to
\begin{equation} \label{precond_eq}
V_1^{-1}M V^{-1}_2 \widehat{x} = \widehat{b}  
\end{equation}
with $\widehat{b} = V_1^{-1}b$ and $\widehat{x} = V_2 x$. The matrices $V_1,V_2$ are 
called the left and right preconditioner, resp., and their product $V = V_1V_2$
is often referred to as {\em the} preconditioner. Note that the spectrum of
$V_1^{-1}MV_2^{-1}$ is identical to that of $V^{-1}M$, so that the effect 
of preconditioning on the eigenvalue distribution is determined by $V$ alone but
not by its factorization $V = V_1V_2$. A preconditioner should approximate      
$M$ (so that the eigenvalues of $V^{-1}M$ cluster around 1). On the other
hand, performing a Krylov subspace method on the preconditioned system requires
multiplies with the preconditioned matrix like in  
$z = V_1^{-1}MV_2^{-1}y$ which are normally 
obtained via
\[
\mbox{solve } V_2w = y, \; v = Mw, \; \mbox{ solve } V_1z = v.
\]
Preconditioning thus introduces additional solves with the matrices $V_1$
and $V_2$ and this overhead should not be too expensive in order to get 
an efficient method. 
A good preconditioner is always a compromise between the latter
requirement and the fact that $V$ should well approximate $M$.

Conceptually, one may distinguish between two types of preconditioners: 
In {\em problem oriented} preconditioners the matrix $V$ is taken as a 
simpler or reduced (with respect to $M$) representation of the underlying
physical problem. {\em Algebraic} preconditioners are obtained directly 
from $M$ without recourse to the application from which $M$ arises. 
Interestingly, algebraic preconditioners seem to be more successful than
problem oriented ones in QCD computations and we therefore focus on 
the latter ones.

\subsection{SSOR preconditioners}  
Each matrix $M$ can be decomposed into
\[
M = D - L - U,
\]
where $D, -L$ and $-U \in \MathC^{n \times n}$ represent the diagonal, the
stricly lower and the stricly upper triangular part of $M$. We assume that 
$M$ has all diagonal elements $\not = 0$, so that $D, D-L$ and $D-U$ are
all non-singular. For a given relaxation parameter $\omega \not = 0$ the SSOR
preconditioner is defined by (see \cite{Sa96}, e.g.)
\[
V = \left(\frac{1}{\omega}D-L\right)D^{-1}\left(\frac{1}{\omega}D-U\right).
\]       
For $\omega = 1$ we thus have $V = M - LD^{-1}U$
as an approximation to $M$. Systems with the preconditioner $V$ are easy to
solve because $\frac{1}{\omega}D-L$ and $\frac{1}{\omega}D-U$ are triangular
so that $x$ in $(\frac{1}{\omega}D - L)x = y$ can be obtained by a simple
forward recursion, and similarly by a backward recursion in $(\frac{1}{\omega}
D - U)x = y$. Note that the situation becomes more involved if we 
consider parallelization issues since recursions are known to parallelize
badly.

Assume that $M$ is of the particular form
\[
M = \left( \begin{array}{cc}
            D_1 & - B_1 \\
            - B_2 & D_2  
           \end{array}
      \right) .
\]
This is the case for the Wilson fermion matrix if we use the standard odd-even
ordering (with $D_1 = D_2 = I$). If we take
$V_1 = (D-L)D^{-1}$ and $V_2 = (D-U)$  we get
\begin{equation} \label{odd_even_red_eq}
V_1^{-1}MV_2^{-1} = \left( \begin{array}{cc}
                             I & 0 \\
                             0 & I - B_2 D_{1}^{-1}B_1 D_2^{-1}
                          \end{array} 
               \right) .
\end{equation}
For the Wilson fermion matrix the second diagonal block in \eqnref{odd_even_red_eq} is 
commonly called the {\em odd-even reduced} system. Our discussion
shows that odd-even reduction is nothing else but the SSOR preconditioning
with respect to the odd-even ordering and with $\omega = 1$. 
Very exceptionally, in this case it is of no harm to calculate
the preconditioned matrix explicitly as done in \eqnref{odd_even_red_eq},
whereas in general this produces too much fill-in to be practicable.
If we re-interprete $D, -L, -U$ as {\em block} parts of $M$, the above
discussion can also be used to derive block SSOR preconditioners.
In QCD this can be useful in the context of improved actions where $D$ then is
block diagonal with blocks of size $12 \times 12$. See also \cite{Gu96,Ja97}

\subsection{ILU factorizations}

The {\em incomplete LU factorization} (ILU) (see \cite{Sa96}, e.g.)
is another algebraic method
to obtain a preconditioner $V = (\widehat{D}- \widehat{L})\widehat{D}^{-1}(\widehat{D} - \widehat{U})$
for $M$ where, again, $\widehat{D}, \widehat{L}, \widehat{U}$ are diagonal, strictly
lower and strictly upper triangular, respectively. These matrices are obtained
by performing a variant of Gaussian elimination on $M$ imposing
restrictions on the
amount of fill-in in the factors $\widehat{D} - \widehat{L}$ and 
$\widehat{D} - \widehat{U}$ so that $V$ represents only an approximate
(incomplete) factorization of $M$. If we allow for no fill-in
(i.e.\ $\widehat{D} - \widehat{L}$ and $\widehat{D} - \widehat{U}$ have the same sparsity 
structure as $M$) and if $M$ represents a nearest neighbor-coupling on 
a regular grid, then $\widehat{L} = L$ and $\widehat{U} = U$, so that the only 
difference to the SSOR preconditioner resides in the diagonal part $\widehat{D}$.
For the Wilson fermion matrix with Wilson parameter $r=1$ both
preconditioners turn out to be totally equal.
ILU preconditioners are often somewhat more efficient than SSOR 
preconditioners, but note that they require a start-up phase to
compute $\widehat{D}$ (and $\widehat{L}$ and $\widehat{U}$, in general).    

\subsection{The Eisenstat trick}
If we have an SSOR or ILU preconditioner of the form $V_1 = 
 (\widehat{D}-L)\widehat{D}^{-1}$ and $V_2 = (\widehat{D}-U)$, the product
$ y = V_{1}^{-1} M V_{2}^{-1} x $ can be computed as
\[
\begin{array}{l}
\mbox{solve } (\widehat{D} - U)v  = x, \\
\mbox{solve } (\widehat{D} - L)w  = (D-2\widehat{D}) v + x,\\
y  =  \widehat{D} (v+w).
\end{array}
\]
As far as flop counts are concerned, the above scheme is as
expensive as one multiplication with $M$ itself, except for some
additional operations involving diagonal matrices which can usually 
be neglected. So, due to the Eisenstat trick, the ILU and SSOR 
preconditioners do not increase the amount of work per iteration,
thus making these preconditioners particularly attractive. Note that 
the Eisenstat trick can also be applied in more general situations, 
see \cite{Ei81}. 

\subsection{The influence of orderings}

When writing down the equation $Mx = b$ we are free to chose any ordering
for the variables, and the change from one ordering to another translates into
a transformation of the kind $M \to P^{\dagger}MP$ with $P$ a permutation 
matrix. For both, the SSOR and ILU preconditioners, the spectrum of the
preconditioned matrix depends on the ordering chosen (but the Eisenstat trick
can always be applied). There is therefore 
a potential to optimize these preconditioners using
the best ordering. Typically, orderings which yield good preconditioners
make the recurrencies in solving the triangular systems less amenable to parallel
implementations. For example, the natural lexicographic ordering of lattice
points in the Wilson fermion matrix was shown to yield a high quality
ILU preconditioner \cite{Oy86}, but it cannot be handled efficiently on a distributed
memory parallel computer. In \cite{Fietal96} it was shown that a new
{\em locally lexicographic}
ordering can yield up to a factor 2 improvement over odd-even preconditioning
on a Quadrics parallel computer. 
 
\subsection{Polynomial preconditioning}

Another algebraic preconditioner is obtained by taking $V^{-1} = s(M)$ 
where $s$ is a polynomial such that $s(M)$ approximates $M^{-1}$. So the
multiplication with $V^{-1}$ requires $\deg(s)$ multiplies with $M$. Consequently,
$deg(s)+1$ steps on the original system are as expensive as 
one step on the preconditioned system
and the iterates are from the same Krylov
subspace. In this respect polynomial 
preconditioning therefore offers little advantage, but it was shown in 
\cite{Gu96} that it can be useful as a mean of stabilizing the MR method
in certain situations.

\section{EXAMPLE: WILSON FERMIONS}

The generic form of the Wilson fermion matrix is $M = I - \kappa B$ where
$B$ represents the nearest neighbor coupling on the space-time lattice. 
Taking the even-odd ordering, $B$ has the form
\[
B  = \left( \begin{array}{cc}
            0 & B_1 \\
            B_2 & 0 
       \end{array} 
     \right).
\]
The odd-even reduced matrix $M_{e}$ from \eqnref{odd_even_red_eq} 
is
\[
M_{e} = I - \kappa^2 \cdot B_2B_1.
\]
A typical example of the eigenvalue distribution for $M$ and $M_{e}$
(calculated from a confined configuration on a
small $4^4$-lattice at $\beta= 5.0$ and $\kappa = 0.150$)
is given on top of Fig.~1. Note that all eigenvalues lie in the right half 
plane so that GMRES($k$) is known to converge for all $k$. A number $\mu$
is an eigenvalue of $M_{e}$ if and only if it is of the form 
$\mu = \lambda(2- \lambda)$ where $\lambda $ is an eigenvalue of $M$.
We write $\alpha_e > 0 $ for the smallest real part of an eigenvalue of
$M_{e}$.   

Both, $M$ and $M_{e}$ are $\gamma_5$-symmetric, and we denote the respective 
symmetrized systems  by
\[
Q = \gamma_5 \cdot M, \enspace Q_{e} = \gamma_5 \cdot M_{e}.
\]
$Q$ and $Q_{e}$ are both hermitian and 
half of their eigenvalues are negative and half are positive.
Moreover, the eigenvalue plots given in Fig.~1 show that except for
a pair close to zero the
eigenvalues are quite evenly distributed in two intervals symmetric to the
origin, denoted $[-b_e,-a_e], [a_e,b_e]$
for $Q_e$.  
Finally, if we consider $M_e^{\dagger}M_e = Q_e^{2}$, then its 
eigenvalues are just the squares of those of $Q_e$ and are therefore distributed
in the interval $[a_e^2,b_e^2]$. 

\setlength{\unitlength}{1in}
\begin{figure}
\begin{picture}(3.4,6.6)(0,0.1)
\put(-0.35,0){ \epsfig{file=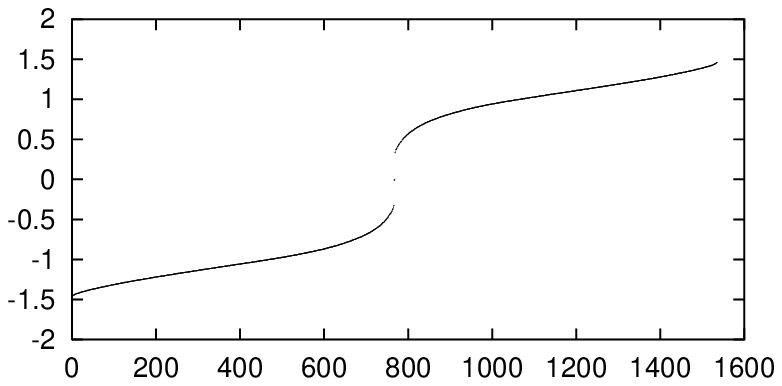,width=3.2in,height=1.5in}}
\put(1.5,0.2){ \epsfig{file=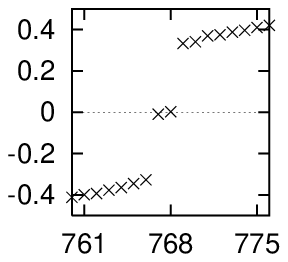,width=1.1in,height=0.8in}}
\put(1.7,-0.1){\makebox(0,0){no.\ of eigenvalue}}
\put(0.4,1.35){\makebox(0,0){$Q_e$}}
\put(-0.35,1.9){ \epsfig{file=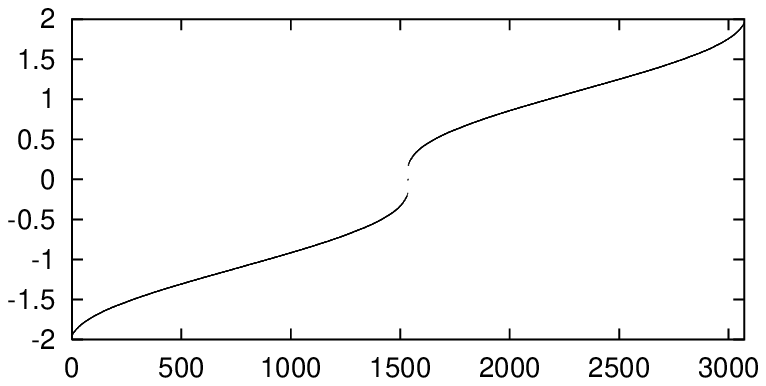,width=3.2in,height=1.5in}}
\put(1.5,2.1){ \epsfig{file=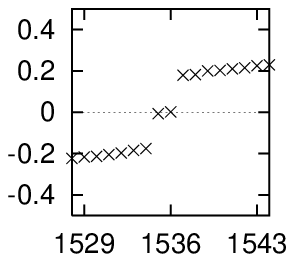,width=1.1in,height=0.8in}}
\put(1.7,1.8){\makebox(0,0){no.\ of eigenvalue}}
\put(0.4,3.25){\makebox(0,0){$Q$}}
\put(-0.35,3.6){ \epsfig{file=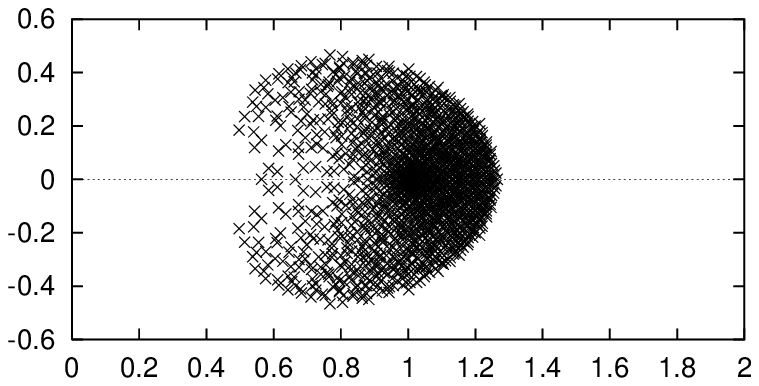,width=3.2in,height=1.5in}}
\put(0.4,4.95){\makebox(0,0){$M_e$}}
\put(-0.35,5.3){ \epsfig{file=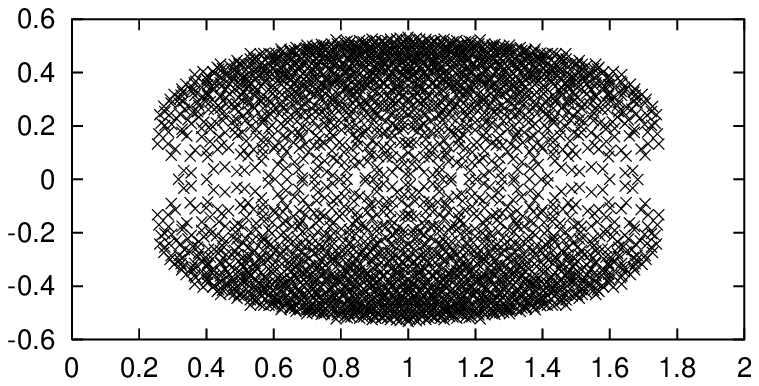,width=3.2in,height=1.5in}}
\put(0.4,6.65){\makebox(0,0){$M$}}
\end{picture}
\caption{Spectra of $M$, $M_e$, $Q$, $Q_e$ for a $4^4$ lattice
at $\beta = 5.0$ and $\kappa = 0.150$. 
The inlays for $Q$, $Q_e$ represent a zoom to the eigenvalues
close to 0.}
\end{figure}

With the information of Fig.~1 as a background, we can now start to 
discuss the condition of the different matrices, i.e. the numbers $c_m$ 
from \eqnref{cm_eq}. First of all we realize that odd-even preconditioning
really makes the spectrum of $M$ and $Q$ `nicer', since eigenvalues are 
mapped away from 0 and are more clustered. We thus restrict the 
subsequent discussion to the even-odd preconditioned matrices.
For the hermitian matrices $Q_e$ and $Q_e^{2}$, rather good bounds for 
$c_m$ can be derived via the Chebyshev polynomials on $[a_e,b_e]$,
and we obtain
\[
c_m(Q_e)  \leq \left( \frac{1- \frac{a_e}{b_e}}{1 + \frac{a_e}{b_e}} 
\right)^{m/2},  
c_m(Q_e^{2})  \leq \left( \frac{1- \frac{a_e}{b_e}}{1 + \frac{a_e}{b_e}}
\right)^m.
\]
Since MINRES is a feasible optimal method for $Q_e$ and CG is an 
optimal method for $Q^2_e= M_e^{\dagger}M_e$, we also can take the above 
numbers as an approximate measure for the performance of these methods. 
They indicate that CGNR, the CG method applied to the normal equations
$M^{\dagger}Mx = M^{\dagger}b$ would require half as many iterations
as MINRES for $Qx = \gamma_5 b$. So in terms of matrix  multiplies
with $Q$ (or $M$) -- which is the computationally dominating part
 --, both methods should be comparable. 
Fig.~2 gives some experimental data,
where we show the convergence history of MINRES and CGNR plotting the
residual norm against the number of matrix mulitplies. We see that 
MINRES actually performs somewhat better than CGNR. The data comes from
a confined configuration on a $16^4$ lattice at $\beta = 6.0$ and 
$\kappa = 0.155$ which yields a relative quark mass of 0.02, approx.      
(In order to observe substantial differences between different
methods it is important to work on `difficult' problems, i.e.\ with 
small relative quark masses.)

For the non-hermitian matrix $M_e$ it is not possible to give an accurate
bound on $c_m$, but we know at least that
\[
c_m(M_e) \leq \left. \left(1 - \frac{\alpha_e^2}{b_e^2} \right)^m \right/ 2 = \rho_m,
\]
and MR (= GMRES(1)) already achieves $\|r^m\| \leq \rho_m \|r^0\|$
\cite{Sa96,EES83}. The remaining parts of Fig.~2
give the convergence history for GMRES($k$) for several values of $k$ 
for the same configuration as before as well as the corresponding
results for BiCG, QMR and BiCGStab. In BiCG and QMR we made use of the
savings due to $\gamma_5$-symmetry. One immediately notices the 
more erratic behavior of BiCG and BiCGStab. We also see that increasing
$k$ in GMRES($k$) gives significant improvement, but there seems little use 
taking  $k$ larger than 8. Finally, QMR, BiCG, BiCGStab  
perform best and 
quite comparably which we can interprete as an indication that they
are all close to optimal for $M_e$. This observation is backed by results from
\cite{Fo95} proving that even the {\em full} GMRES method did not
give substantial improvement over BiCG, QMR or BiCGStab.

\subsection*{Acknowledgement}
I am grateful for the continuing 
excellent cooperation
with K.~Schilling's group at Wuppertal and J\"ulich, particularly
to Th.~Lippert. The numerical results were obtained 
together with P.~Fiebach from the Department
of Mathematics in Wuppertal.       

\setlength{\unitlength}{1in}
\begin{figure}[tbh]
\begin{picture}(3.4,5.0)(0,0.2)
\put(-0.35,0){ \epsfig{file=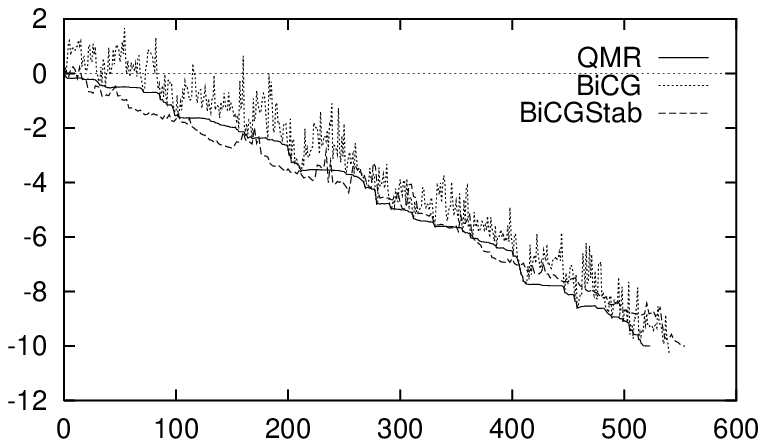,width=3.2in,height=1.5in}}
\put(0.65,0.8){\makebox(0,0){$\log_{10}(\frac{\|r^m\|}{\|r^0\|})$}}
\put(1.7,-0.1){\makebox(0,0){$\#$ matrix mulitplies}}
\put(-0.35,1.8){ \epsfig{file=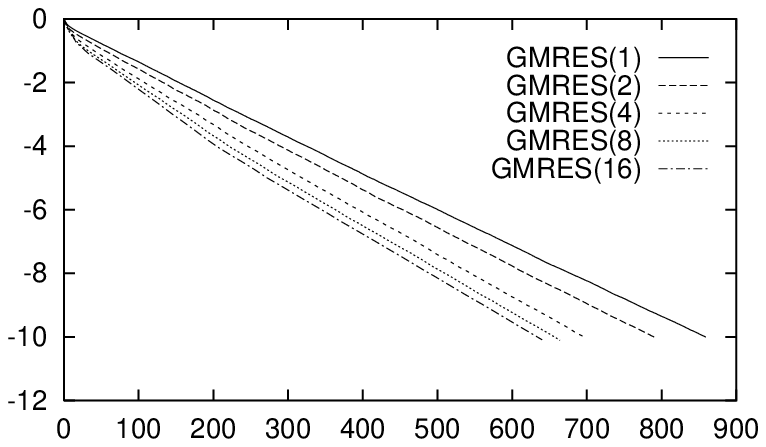,width=3.2in,height=1.5in}}
\put(0.65,2.6){\makebox(0,0){$\log_{10}(\frac{\|r^m\|}{\|r^0\|})$}}
\put(1.7,1.7){\makebox(0,0){$\#$ matrix mulitplies}}
\put(-0.35,3.6){ \epsfig{file=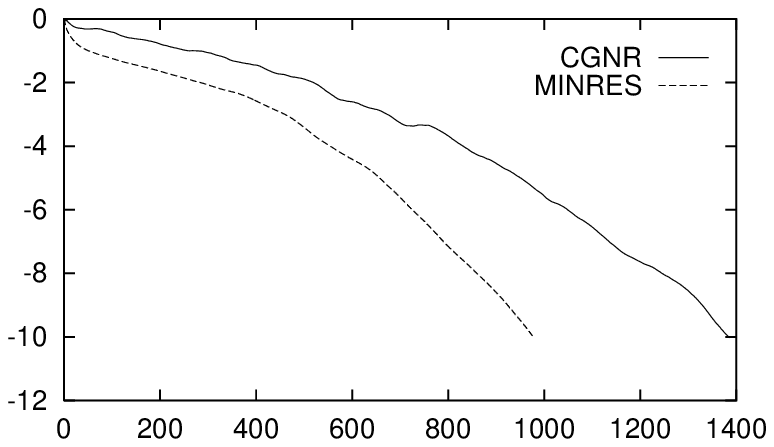,width=3.2in,height=1.5in}}
\put(0.65,4.4){\makebox(0,0){$\log_{10}(\frac{\|r^m\|}{\|r^0\|})$}}
\put(1.7,3.5){\makebox(0,0){$\#$ matrix mulitplies}}
\end{picture}
\caption{Convergence histories for different methods.}
\end{figure}

\end{document}